\documentclass[aps,prc,a4paper,superscriptaddress]{revtex4-1}
\usepackage[dvipdfmx]{graphicx}
\usepackage{bm}
\usepackage{cancel}
\usepackage{color}
\usepackage{multirow}
\usepackage{slashbox}
\usepackage{amsmath}
\usepackage{amssymb}

\begin{document}

\title{Resonance in the $Y_{c}N$ potential model}
\date{\today}

\author{Saori Maeda}
\email[E-mail: ]{smaeda.nucl.phys@gmail.com}
\affiliation{Department of Physics, Tokyo Institute of Technology, O-okayama 2-12-1, Meguro, Tokyo 152-8551, Japan}
\author{Makoto Oka}
\affiliation{Department of Physics, Tokyo Institute of Technology, O-okayama 2-12-1, Meguro, Tokyo 152-8551, Japan}
\affiliation{Advanced Science Research Center, Japan Atomic Energy Agency, Tokai, Ibaraki 319-1195, Japan}
\author{Yan-Rui Liu}
\affiliation{Department of Physics, Shandong University, Jinan, Shandong 250100, China}

\begin{abstract}

We calculate two-body $J^{\pi}=0^{+}, 1^{+}$, and $J^{\pi}=2^{+}$ resonance states 
of $Y_{c}$ ($= \Lambda_{c}$, $\Sigma_{c}$, or $\Sigma_{c}^{*}$) and $N$ using the complex scaling method.
We employ the $Y_{c}N$-CTNN potentials, which were proposed in our previous study, 
and obtain four resonances near $\Sigma_{c}N$ and $\Sigma_{c}^{*}N$ thresholds.
From the analysis by the binding energies of partial channel systems, we conclude that these resonance states are Feshbach resonances.
We compare the results with the $Y_{c}N$ resonance states in the heavy quark limit, where the $\Sigma_{c}N$ and $\Sigma_{c}^{*}N$ thresholds are degenerate, 
and find that they form two pairs of the heavy-quark doublets in agreement with the heavy quark spin symmetry.

\end{abstract}

\maketitle

\section{Introduction}
\label{sec:res2-1}

Recent discoveries of new and exotic charmed hadrons have stimulated the study of charmed hadrons 
both from experimental and theoretical viewpoints.
An interesting and not-yet-well-explored subject is the charmed nucleus, {\it i.e.}, a bound state of charmed baryon(s) in nuclei.
As the charm quark is much heavier than the strange quark and thus the interaction of charm has novel properties and symmetry,
it is intriguing to explore the interactions and properties of charmed baryons in nuclear medium.
Let us consider the $\Lambda_c$-nuclear bound state. 
The system may look like an ordinary hypernucleus with the $\Lambda$ hyperon replaced by $\Lambda_c$. 
It is, however, different from a few important aspects.
Because $\Lambda_c N$ couples strongly to $\Sigma_c N$ and $\Sigma_c^* N$ systems, the contribution from
the $\Sigma_c$ sector becomes significant. 
Furthermore, the heavy-quark spin symmetry of QCD predicts that the spin-dependent interaction of charm
is much weaker than strangeness, and thus the spin dependent
$\Lambda_c N$ force is negligible.

In a previous paper\cite{ref-sm}, we studied the potential model of the $Y_{c}N$ 2-body system, 
where $Y_c$ denotes the charmed baryon, $\Lambda_{c}$, $\Sigma_{c}$, or $\Sigma_{c}^{*}$.
We considered the $J^\pi=0^+$ and $1^+$ states of $\Lambda_c N$ with full couplings to the corresponding
$\Sigma_c N$ and $\Sigma_c^* N$ channels.
The diagonal and off-diagonal potentials among the channels are calculated in the meson exchange
picture, supplemented by the short-range repulsion taken from the quark exchange model.
The $D$ wave mixings due to the tensor force of the pion exchange potential are taken into account.
We pointed out that a shallow $\Lambda_c N$ bound state may exist both in $J^\pi=0^+$ and $1^+$,
where mixings of $\Sigma_c N$ and $\Sigma_c^* N$ play significant roles.
The near-degeneracy of $J^\pi=0^+$ and $1^+$ states indicates the heavy-quark spin symmetry 
in the two-baryon system.

In this paper, we extend our analysis at above the $\Lambda_c N$ threshold and study resonance
states around the $\Sigma_c N$ and $\Sigma_c^* N$ thresholds.
It happens  that there exist bound states in the $\Sigma_c N$ (or $\Sigma_c^* N$) single channel 
potential, if we omit the coupling to the $\Lambda_c N$ channel.
Then a corresponding sharp (Feshbach) resonance may appear below the $\Sigma_c N$ (or $\Sigma_c^* N$)
threshold.
We apply the complex scaling method to the coupled channel potential and obtain the resonance energy and width.

In the literature, the $Y_{c}N$ interaction has been studied in various approaches, such as
the constituent quark model \cite{ref-02,ref-022} and the lattice QCD\cite{ref-hal1,ref-hal2, ref-lat}.
In particular, the references \cite{ref-02,ref-hal2} also suggest existence of resonance states.

In section \ref{sec:res2-2}, we review the $Y_{c}N$ potentials in our previous study,
in which four variants of the potentials were proposed, and summarize their properties.
In section \ref{sec:res-2-3}, we give the formulation of the complex scaling method.
Results of the calculation are presented in section \ref{sec:res2-4}.
We discuss the properties of the obtained resonances in section \ref{sec:res2-5}, 
where we interpret the resonance states from the viewpoint of heavy-quark spin doublet.
A summary is given in section \ref{sec:res2-6}.

\section{Formulation}

We study the $Y_{c}N$ two-body system with $J^{\pi}=0^{+}$, $1^{+}$, and $2^{+}$ by solving the Schr\"odinger equation, 
\begin{eqnarray}
(H^{J} - E^{J}) \Psi^{J} = 0, \nonumber \\
\Psi^{J} = \left(
\begin{array}{c} \Psi^{J}_{1} \\
\vdots \\
\Psi^{J}_{\beta}
\end{array} \right),
\end{eqnarray}
where $\beta$ is the label of the coupled channels.
Tables \ref{tb:res2-1} and \ref{tb:res2-2} show possible channels contributing in $\Lambda_{c}N$, $\Sigma_{c}N$, and $\Sigma_{c}^{*}N$.

\begin{table}[t]
\centering
\small
\begin{tabular}{|l|ccccccc|} \hline
Channels $\beta$ & 1 & 2 & 3 & 4 & 5 & 6 & 7 \\ \hline
$J^{\pi} = 0^{+}$ & $\Lambda_{c}N({}^{1}S_{0})$ & $\Sigma_{c}N({}^{1}S_{0})$ & $\Sigma_{c}^{*}N({}^{5}D_{0})$ &   &   &   &   \\
$J^{\pi} = 1^{+}$ & $\Lambda_{c}N({}^{3}S_{1})$ & $\Sigma_{c}N({}^{3}S_{1})$ & $\Sigma_{c}^{*}N({}^{3}S_{1})$ & $\Lambda_{c}N({}^{3}D_{1})$ & $\Sigma_{c}N({}^{3}D_{1})$ & $\Sigma_{c}^{*}N({}^{3}D_{1})$ & $\Sigma_{c}^{*}N({}^{5}D_{1})$ \\ \hline
\end{tabular}
\caption{The S-wave $\Lambda_{c}N$ states and the channels coupling to them}
\label{tb:res2-1}
\end{table}

\begin{table}[t]
\centering
\small
\begin{tabular}{|l|cccc|} \hline
Channels $\beta$ & 1 & 2 & 3 & 4 \\ \hline
\multirow{3}{*}{$J^{\pi} = 2^{+}$} & $\Lambda_{c}N({}^{1}D_{2})$ & $\Sigma_{c}N({}^{1}D_{2})$ & $\Lambda_{c}N({}^{3}D_{2})$ & $\Sigma_{c}N({}^{3}D_{2})$ \\ \cline{2-5}
 & 5 & 6 & 7 & 8 \\ \cline{2-5}
  & $\Sigma_{c}^{*}N({}^{3}D_{2})$ & $\Sigma_{c}^{*}N({}^{5}S_{2})$ & $\Sigma_{c}^{*}N({}^{5}D_{2})$ & $\Sigma_{c}^{*}N({}^{5}G_{2})$  \\ \hline
\end{tabular}
\caption{The D-wave $\Lambda_{c}N$ states and the channels coupling to them}
\label{tb:res2-2}
\end{table}

In the two-body $Y_{c}N$ binding energy calculation, $J^{\pi}=2^{+}$ state is not taken into account 
because there is no bound $2^{+}$ state.
On the other hand, the $J^{\pi}=2^{+}$ states include $\Sigma_{c}^{*}N$ S-wave channel.
The $\Sigma_{c}^{*}$ channels are characteristic of the two-body system including a charmed baryon.
To consider the effects of these channels, 
we study the $J^{\pi}=2^{+}$ state which seems to have resonance states from the coupling with the $\Sigma_{c}^{*}N$ channel.

\subsection{Hamiltonian}
\label{sec:res2-2}

In this study, we introduce the $Y_{c}N$-CTNN potentials between a charmed baryon and a nucleon 
obeying the heavy quark symmetry, chiral symmetry, and hidden local symmetry \cite{ref-had1,ref-had2,ref-had3,ref-had4}.
The meson exchange part contains four kinds of terms: spin-independent, spin-spin, spin-orbit, and tensor,  
\begin{eqnarray}
H^{J} & = & T + V_{Y_{c}N} \nonumber \\
 & = & T + V_{\pi}(i,j) +V_{\sigma}(i,j) + V_{QCM} \nonumber
\end{eqnarray}
\begin{eqnarray}
V_{\pi}(i,j) & = & C_{\pi}(i,j)\frac{m_{\pi}^{3}}{24\pi f_{\pi}^{2}}\left\{ \left< \bm{\mathcal{O}}_{spin} \right>_{ij} Y_{1}(m_{\pi}, \Lambda_{\pi} ,r) + \left< \bm{\mathcal{O}}_{ten} \right>_{ij} H_{3}(m_{\pi}, \Lambda_{\pi} ,r) \right\} \nonumber \\
V_{\sigma}(i,j) & = & C_{\sigma}(i,j)\frac{m_{\sigma}}{16\pi}\left\{ \left< \bm{1} \right>_{ij}4Y_{1}(m_{\sigma}, \Lambda_{\sigma} ,r) + \left< \bm{\mathcal{O}}_{LS} \right>_{ij} \left( \frac{m_{\sigma}}{M_{N}} \right)^{2} Z_{3}(m_{\sigma}, \Lambda_{\sigma} ,r) \right\},
\label{eq:res2-1}
\end{eqnarray}
where the operator matrix elements are given in \cite{ref-sm} with the exception of the $J^{\pi} = 2^{+}$ state.
The operator matrix elements for $J^{\pi} = 2^{+}$ are shown in Appendix \ref{app:obep2}.
The quark model repulsive potential is defined by the Gaussian function, 
\begin{equation}
V_{QCM}=V_{0}e^{-(r^{2}/b^{2})}, \nonumber
\label{eq:res2-2}
\end{equation}
where the parameter $V_{0}$ given in \cite{ref-2}.

In this calculation, we adopt mainly the CTNN-d parameter, which is the most attractive parameter set among the four parameter sets.
The parameters are given in eq.(\ref{eq:res2-3}) and Table \ref{tb:res2-0}, 
\begin{eqnarray}
\Lambda_{\pi} & = & 750.0{\rm [MeV]}, \nonumber \\
\Lambda_{\sigma} & = & 1000.0{\rm [MeV]}, 
\label{eq:res2-3}
\end{eqnarray}
\begin{table}[h]
\centering
\begin{tabular}{c|c||ccc} \hline
 & {\rm d} & {\rm a} & {\rm b} & {\rm c} \\ \hline
b{\rm [fm]} & 0.5 & 0.6 & 0.6 & 0.5 \\ 
$C_{\sigma}$ & -70.68 & -67.58 & -77.5 & -60.76 \\ \hline
\end{tabular}
\smallskip
\caption{The b and $C_{\sigma}$ of $Y_{c}N$-CTNN potential for each parameter set}
\label{tb:res2-0}
\end{table}

\subsection{Complex scaling method}
\label{sec:res-2-3}

To investigate the $Y_{c}N$ resonance states, we use the Complex Scaling Method (CSM) \cite{ref-csm1,ref-csm2,ref-csm3,ref-csm4}.
This method introduces the complex parameter {\it i.e.}, a scaling parameter, which changes the Hamiltonian, together with  wave functions and eigenvalues, into complex numbers.
The method gives the resonance energy in real part and the width in imaginary part simultaneously.
In this method we can discuss the properties of the resonance state as an extension of the variational calculation for bound states, though the calculation with this method becomes complicated due to the treatment of complex eigenvalue problem.
We employ the Gaussian Expansion Method (GEM) \cite{ref-6} and calculate the Schr\"odinger equation with the complex energy eigenvalues.

In the CSM, we rotate the radial coordinate, $r$ and the momentum, $k$, to complex numbers on the complex plane.
The transform operator $U(\theta)$ is defined as, 
\begin{equation}
U(\theta) : \left\{ 
\begin{array}{l}
    r \rightarrow r e^{i\theta} \\
    k \rightarrow k e^{-i\theta}
  \end{array} \right.
\end{equation}
where the rotation angle $\theta$ is a real parameter.
We apply this rotation to the Schr\"odinger equation with complex Hamiltonian.
So the Schr\"odinger equation can be written as follows, 
\begin{eqnarray}
{\mathcal H}(r) \Psi = E \Psi \xrightarrow[]{U(\theta)} {\mathcal H}(e^{i\theta}r) \Psi_{\theta} = E_{\theta} \Psi_{\theta} \nonumber \\
{\mathcal H}(e^{i\theta}r) = U(\theta) {\mathcal H}(r) U(\theta)^{-1}, \ \ \ \ \ \Psi_{\theta} = U(\theta)\Psi
\end{eqnarray}
where $\Psi(r)$ is basis function of variational method and 
${\mathcal H}(r)$ is a Hamiltonian,and ${\mathcal H}(r,\theta)$ is transformed Hamiltonian, written by $r$ and $\theta$,
\begin{eqnarray}
{\mathcal H}(r) &=& T + V \nonumber \\ 
T &=& -\frac{1}{2m}(\frac{\partial^{2}}{\partial r^{2}} + \frac{2}{r}\frac{\partial}{\partial r}) \rightarrow -\frac{1}{2m} e^{-i 2 \theta}(\frac{\partial^{2}}{\partial r^{2}} + \frac{2}{r}\frac{\partial}{\partial r}) \nonumber \\
\end{eqnarray}
where T is the kinetic term and V is the potential term.

\begin{figure}[!t]
\centering
\includegraphics[width=100mm]{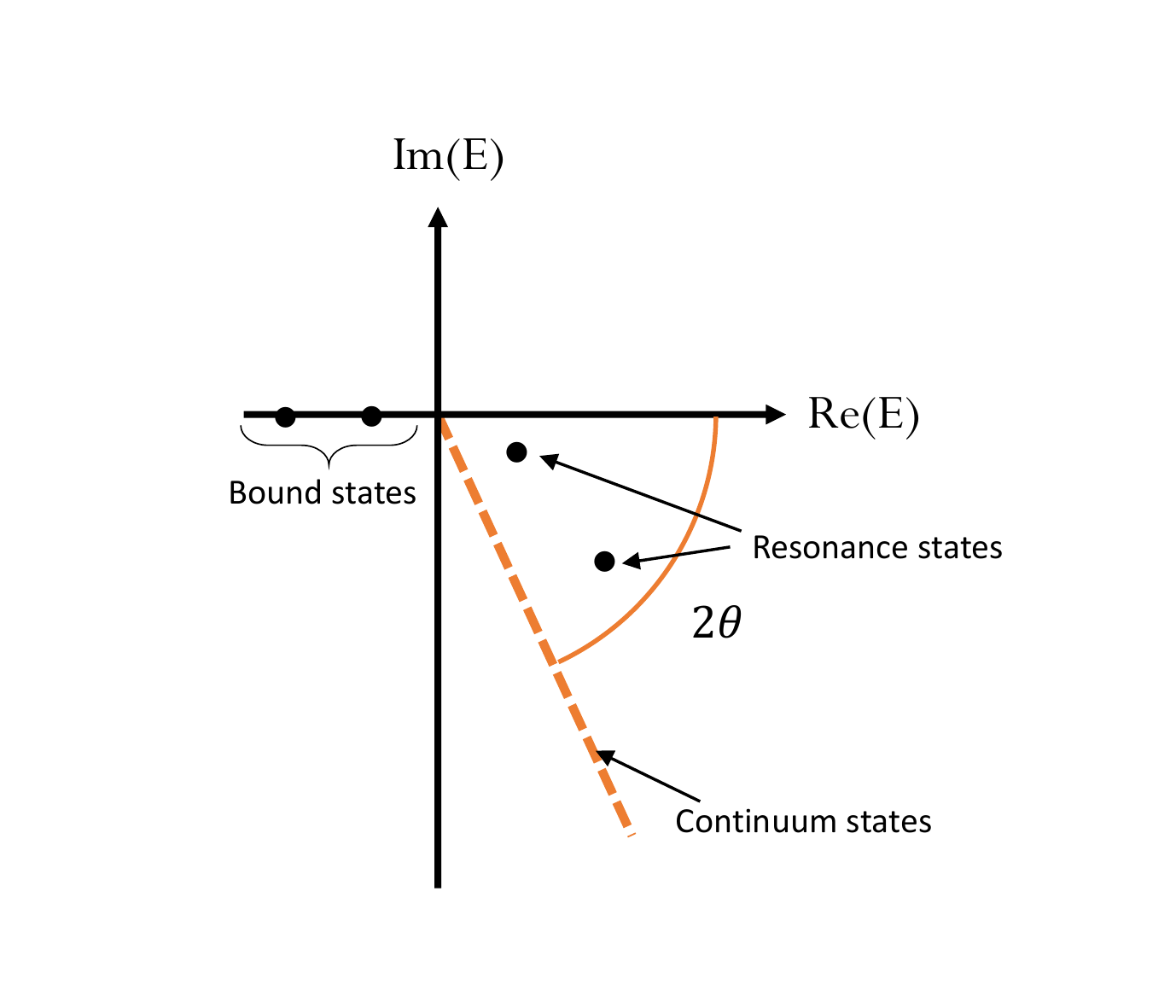}
\caption{The resolution of Identity in Complex Scaling Method}
\label{gr:csm-2}
\end{figure}

Now we consider the behavior of bound states, resonance states, and continuum states with CSM as illustrated in Fig. \ref{gr:csm-2}.
For the coupled channel calculation, the origin means the mass of the threshold of the considered channel.
When the transformation by the complex scaling is applied to $\Psi_{resonance}(r)$, 
it changes to $\Psi_{\rm{resonance}}(re^{i\theta})$, whose asymptotic behavior is given by
\begin{eqnarray}
\Psi_{\rm{resonance}}(re^{i\theta}) \xrightarrow[r \to \infty]{} \exp \{ i(\kappa - i\gamma)(\cos \theta + i \sin \theta)r \} \nonumber \\
=\exp \{ -(\kappa \sin \theta - \gamma \cos \theta)r + i(\kappa \cos \theta + \gamma \sin \theta) r \}.
\end{eqnarray}
This form of wave function is convergent at large $r$ with $\tan \theta > \gamma / \kappa$ or $\tan ( 2\theta ) > \Gamma / (2 E_{r})$.
So the resonance state satisfies the condition that the wave function becomes zero for $r \to \infty$ 
and a resonance energy can be calculated by using compact (Gaussian) wave functions.
The complex energy for the resonance is expressed by  
\begin{equation}
E_{\rm{resonance}}= E_{r} - i \frac{1}{2}\Gamma,
\end{equation}
where $E_{r}$ is the energy of resonance states (resonance energy), 
and $\Gamma$ is the width of the resonance.
Thus, we can calculate the resonance energy and the decay width simultaneously.

\section{Results}
\label{sec:res2-4}

\begin{table}[h]
\centering
\begin{tabular}{c|c|cc} \hline
 \multicolumn{2}{c|}{states} & \multicolumn{2}{c}{B.E. [MeV]}  \\ \hline
threshold & $J^{\pi}$ & from $\Lambda_{c}N$ & from $\Sigma_{c}N$ ($\Sigma_{c}^{*}N$)  \\ \hline
\multirow{2}{*}{$\Sigma_{c}N$} & $J^{\pi}=0^{+}$ & 164.9 & (-2.2)  \\
 & $J^{\pi}=1^{+}$ & 152.8 & (-14.3)  \\ \hline
\multirow{2}{*}{$\Sigma_{c}^{*}N$} & $J^{\pi}=1^{+}$ & 223.8 & (-7.7)  \\
 & $J^{\pi}=2^{+}$ & 213.9 & (-17.6) \\ \hline
\end{tabular}
\smallskip
\caption{The binding energy (B.E.) and resonance state with $Y_{c}N$-CTNN potential d-parameter}
\label{tb:res2-0}
\end{table}

The results of bound state problem for partial channels considerations are shown in Table \ref{tb:res2-0}.
These binding energies are larger than those of $Y_{c}N$ full channel calculation.
In the case of $J^{\pi} = 0^{+}$, there is bound states for $\Sigma_{c}N$ channel, 
even though $\Sigma_{c}N$ is a single channel without non-diagonal term. 
The binding energy is larger than that of full channel calculation. 
In the $J^{\pi} = 1^{+}$ case, there are bound states in both $\Sigma_{c}N$ and $\Sigma_{c}^{*}N$ channels.
We find that the binding energy of $\Sigma_{c}N$ with $J^{\pi} = 1^{+}$ is larger than that in the $J^{\pi} = 0^{+}$ case.
On the other hand, the result of $\Sigma_{c}^{*}N$ channels has almost the half value of $\Sigma_{c}N$ calculation.
In $J^{\pi} = 2^{+}$, there is the large binding energy for $\Sigma_{c}^{*}N$ only.
For $\Lambda_{c}N$ and $\Sigma_{c}N$ channels, there is no bound state due to the high ($L = 2$) angular momentum.
With the same reason, there is no $\Sigma_{c}^{*}N$ bound states in $J^{\pi} = 0^{+}$.

\begin{figure}[!h]
\begin{tabular}{cc}
\begin{minipage}{0.5\hsize}
\begin{center}
\includegraphics[width=80mm]{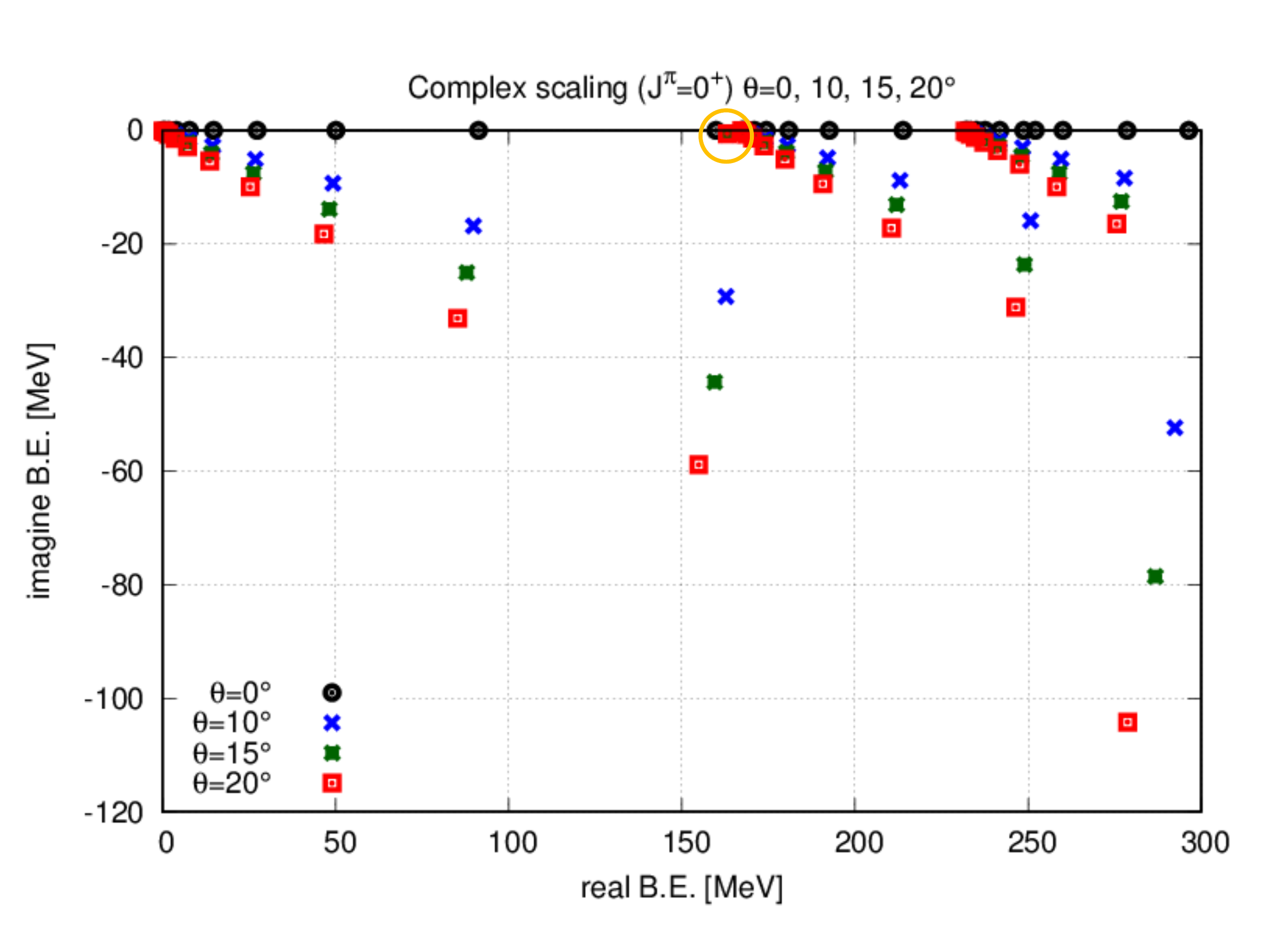}
\caption{\footnotesize{The result of complex scale method in complex plane for $J^{\pi}=0^{+}$}}
\label{gr:res2-1}
\end{center}
\end{minipage}
\begin{minipage}{0.5\hsize}
\begin{center}
\includegraphics[width=80mm]{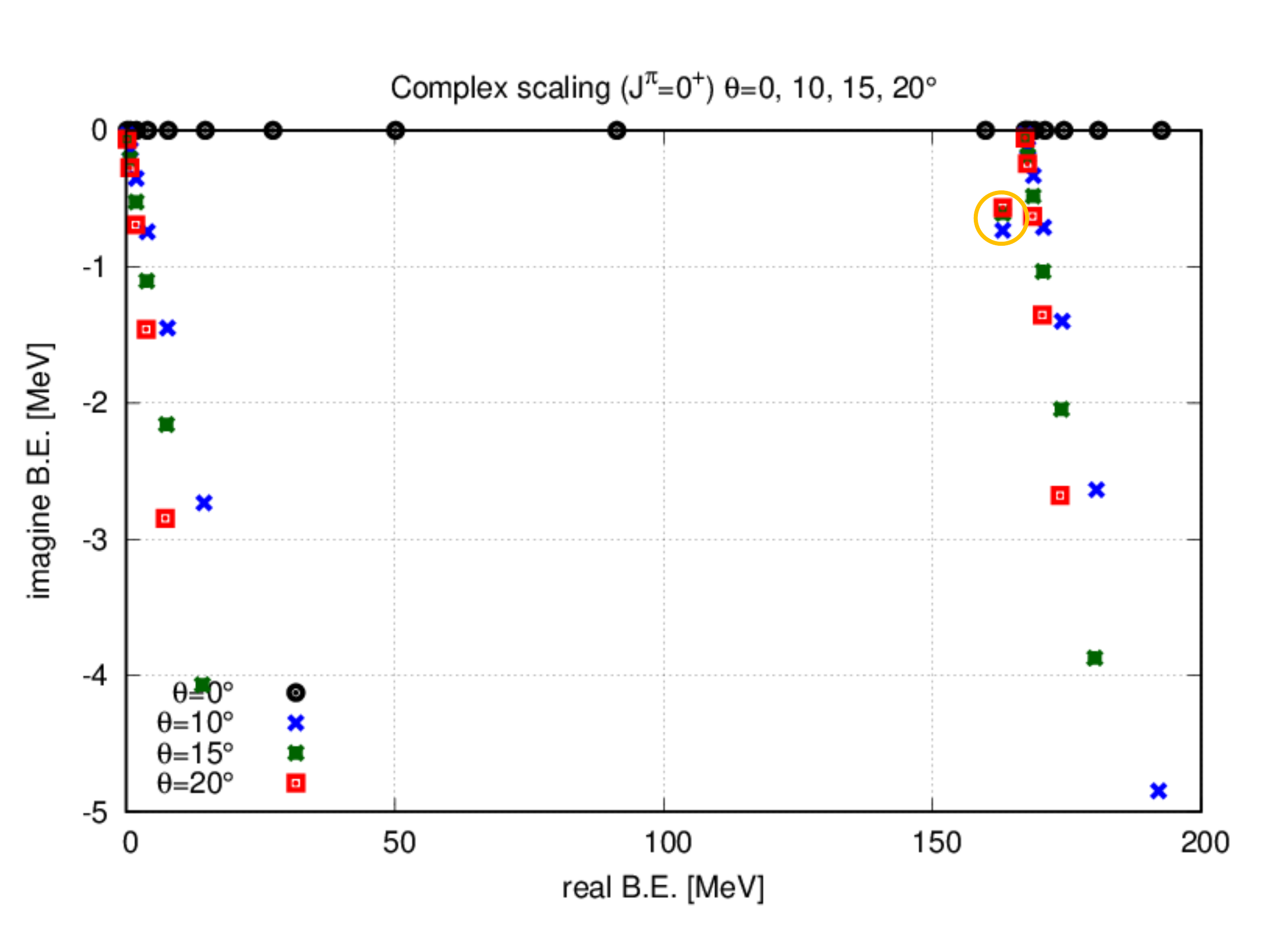}
\caption{\footnotesize{The enlarged graph of Fig. \ref{gr:res2-1} focused near the resonance state.}}
\label{gr:res2-2}
\end{center}
\end{minipage}
\end{tabular}
\end{figure}

\begin{figure}[!h]
\begin{tabular}{cc}
\begin{minipage}{0.5\hsize}
\begin{center}
\includegraphics[width=80mm]{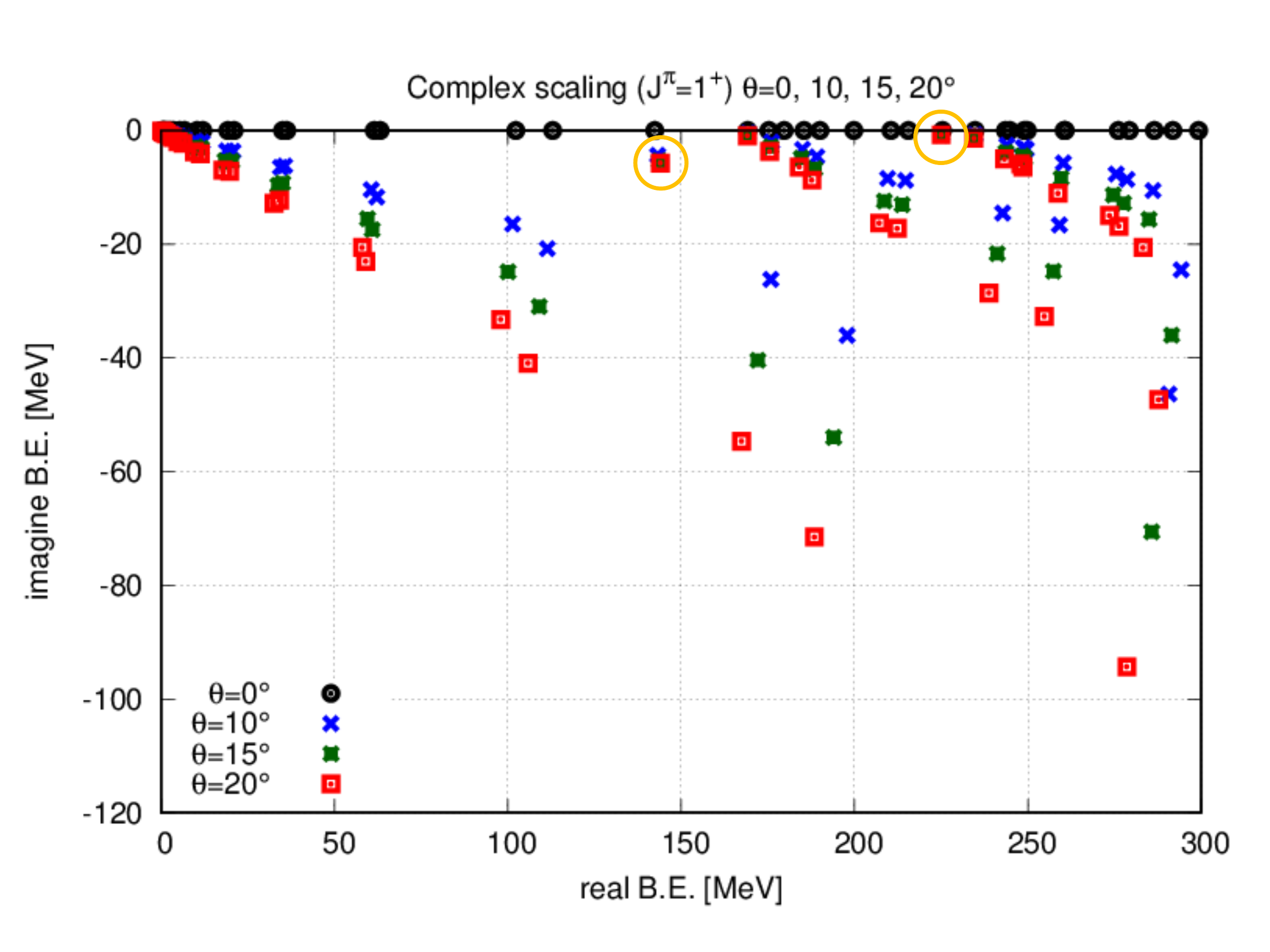}
\caption{\footnotesize{The result of complex scale method in complex plane for $J^{\pi}=1^{+}$}}
\label{gr:res2-3}
\end{center}
\end{minipage}
\begin{minipage}{0.5\hsize}
\begin{center}
\includegraphics[width=80mm]{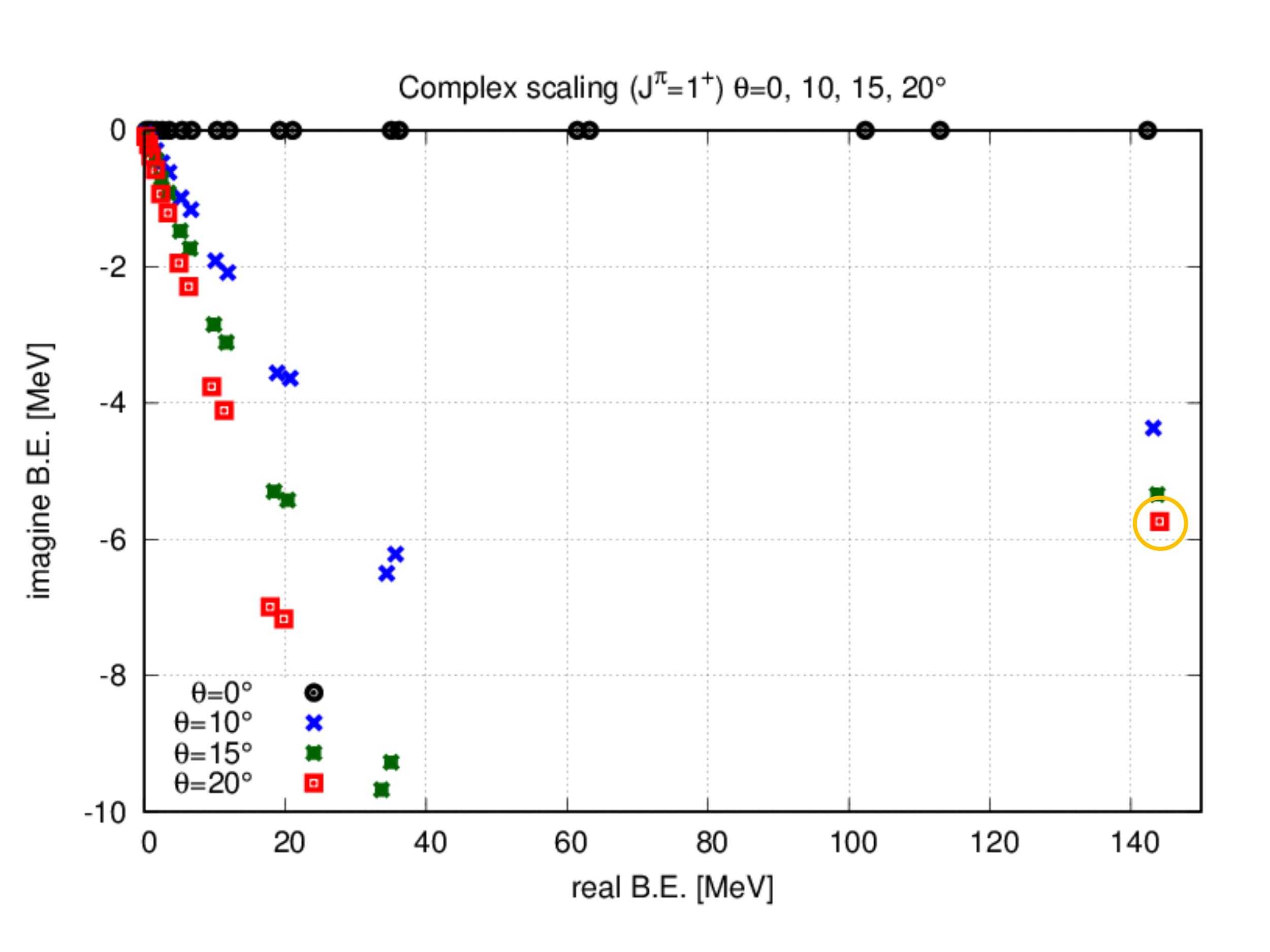}
\caption{\footnotesize{The enlarged graph of Fig. \ref{gr:res2-3} focused near the $\Sigma_{c}N$ resonance state.}}
\label{gr:res2-4}
\end{center}
\end{minipage}
\end{tabular}
\end{figure}

\begin{figure}[!h]
\begin{center}
\includegraphics[width=80mm]{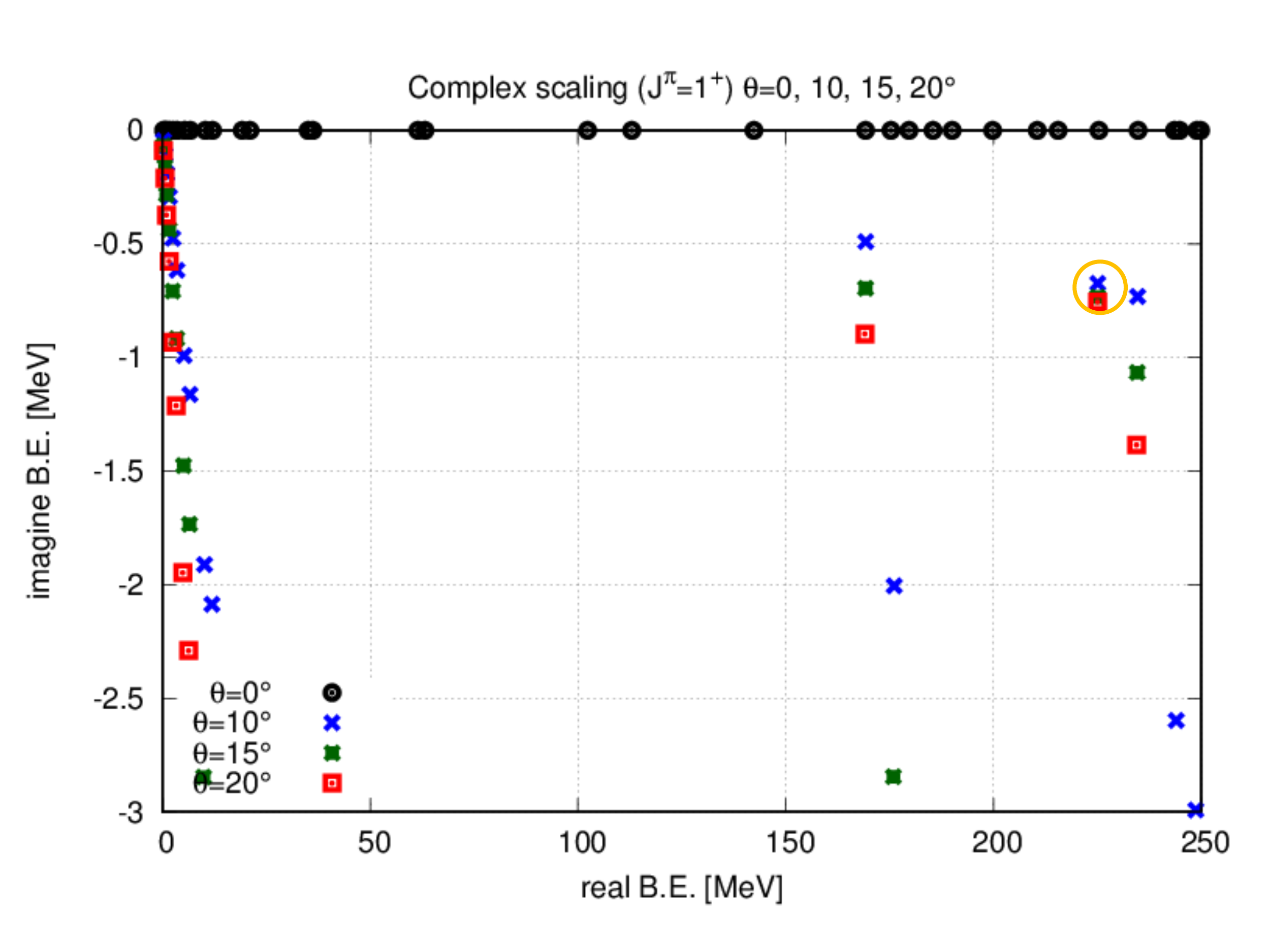}
\caption{\footnotesize{The enlarged graph of Fig. \ref{gr:res2-3} focused near the $\Sigma_{c}^{*}N$ resonance state.}}
\label{gr:res2-5}
\end{center}
\end{figure}

\begin{figure}[!t]
\begin{tabular}{cc}
\begin{minipage}{0.5\hsize}
\begin{center}
\includegraphics[width=80mm]{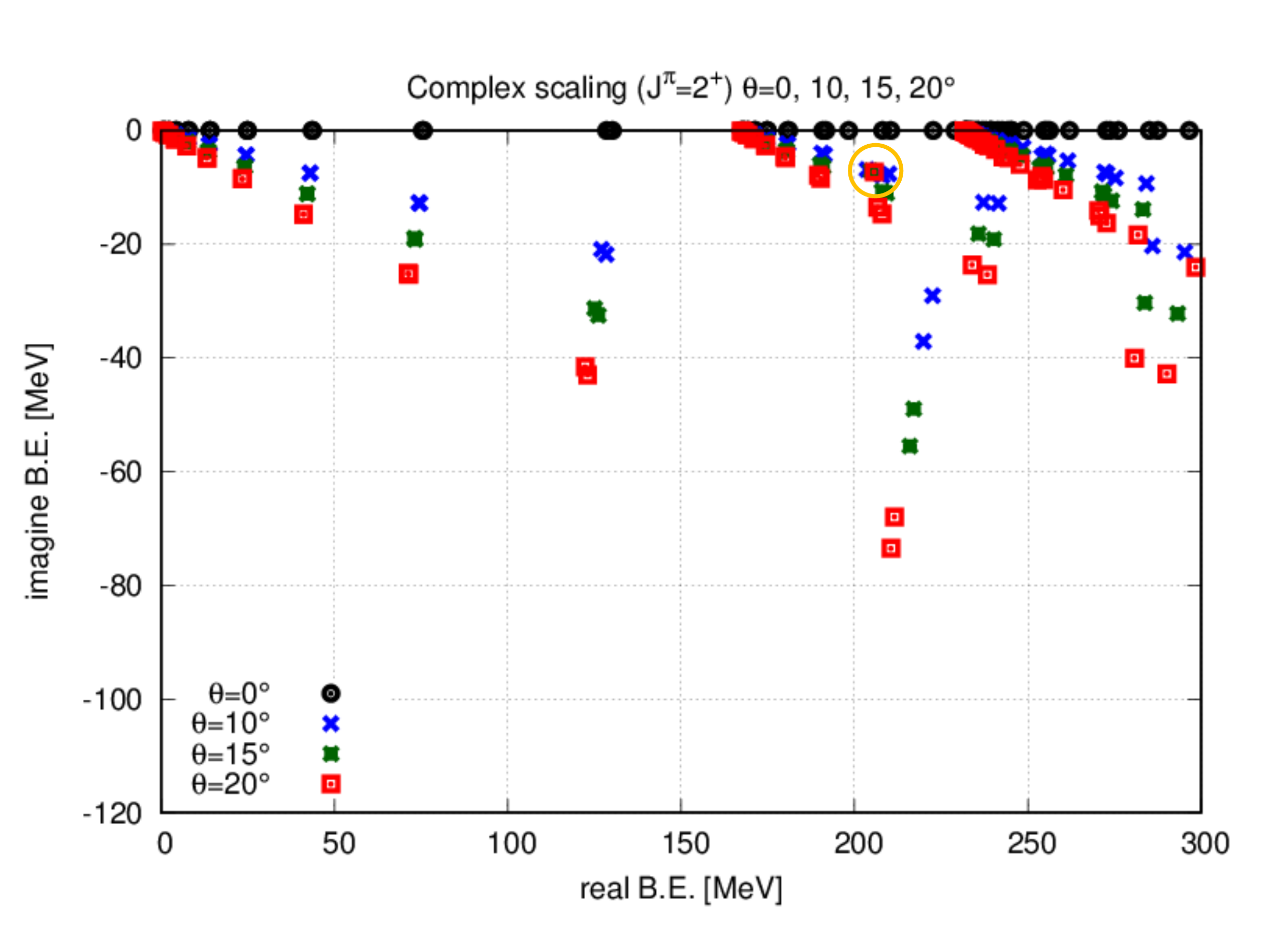}
\caption{\footnotesize{The result of complex scale method in complex plane for $J^{\pi}=2^{+}$}}
\label{gr:res2-6}
\end{center}
\end{minipage}
\begin{minipage}{0.5\hsize}
\begin{center}
\includegraphics[width=80mm]{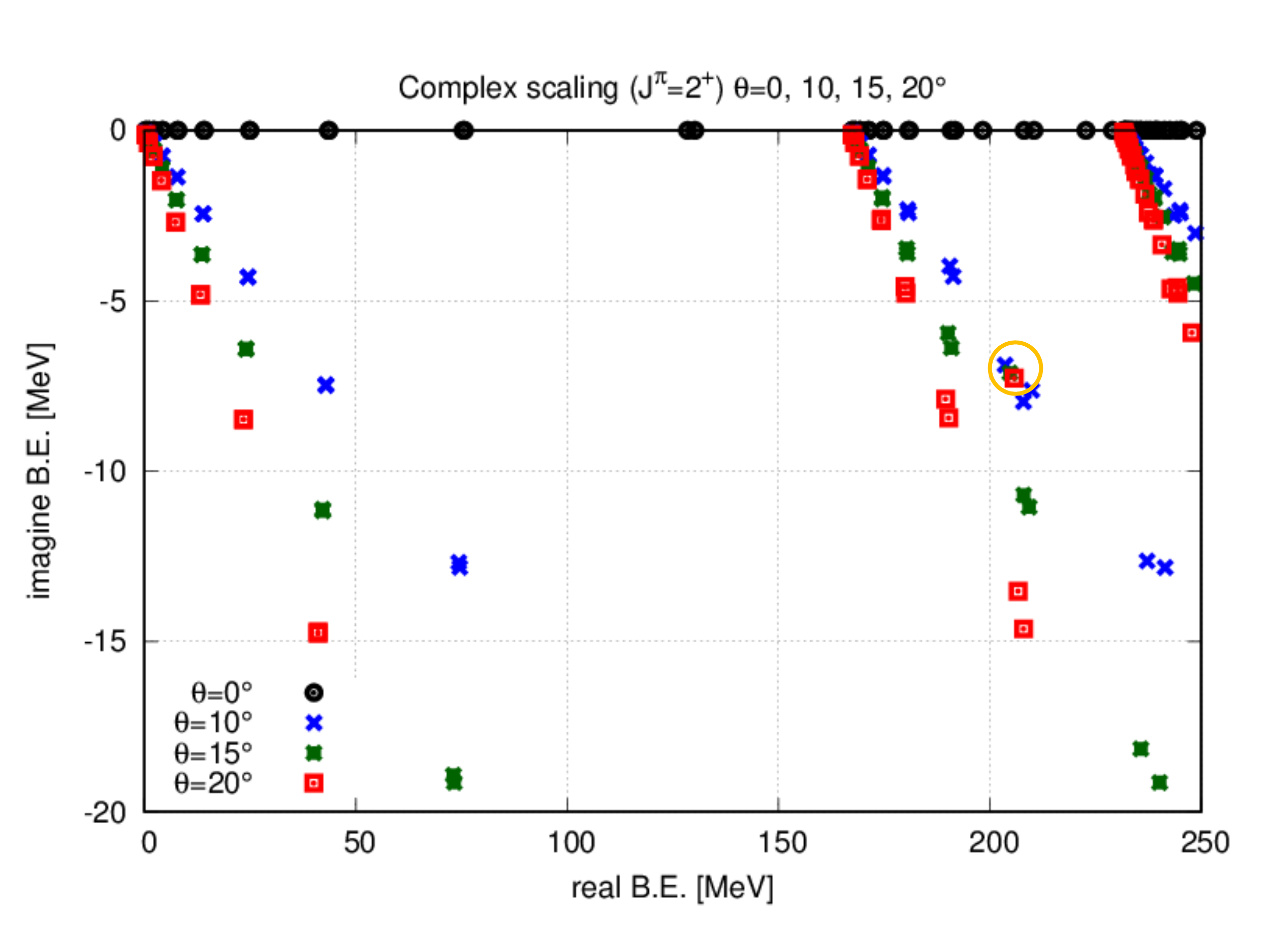}
\caption{\footnotesize{The enlarged graph of Fig. \ref{gr:res2-5} focused near the resonance state.}}
\label{gr:res2-7}
\end{center}
\end{minipage}
\end{tabular}
\end{figure}

Figures \ref{gr:res2-1} and \ref{gr:res2-2} show the CSM results for the $J^{\pi}=0^{+}$ states in the coupled channel calculation.
The complex eigenvalues of the Hamiltonian are plotted for several values of the complex scaling parameter $\theta=10^\circ, 15^\circ$, and $20^\circ$.
When the angle $\theta$ changes, most of eigenenergies move along the threshold lines.
It is, however, noticed that one eigenvalue just below to the $\Sigma_{c}N$ threshold does not move in the order of 1 MeV when $\theta$ changes.
This result indicates a resonance states.
Similar calculations for $J^{\pi}=1^{+}$ and $J^{\pi}=2^{+}$ are shown in Figs. \ref{gr:res2-3} - \ref{gr:res2-6}.


\begin{table}[h]
\centering
\begin{tabular}{c|cc|c} \hline
  & \multicolumn{2}{c|}{Resonance energy [MeV]} &\multirow{2}{*}{width[MeV]} \\
 states & numerical result & (from threshold) &  \\ \hline
 $J^{\pi}=0^{+}$ ($\Sigma_{c}N$) & 163 & (-4) & 1 \\
 $J^{\pi}=1^{+}$ ($\Sigma_{c}N$) & 144 & (-23) & 12 \\ \hline
 $J^{\pi}=1^{+}$ ($\Sigma_{c}N^{*}$) & 225 & (-7) & 2 \\
 $J^{\pi}=2^{+}$ ($\Sigma_{c}N^{*}$) & 206 & (-25) & 14 \\ \hline
\end{tabular}
\smallskip
\caption{Numerical results of the resonance states}
\label{tb:res2-3}
\end{table}

The results are summarized in Table \ref{tb:res2-3}.

In the $J^{\pi}=0^{+}$ channel, there is one resonance state near the $\Sigma_{c}N$ threshold.
The state is very close to the $\Sigma_{c}N$ threshold and has a narrow width.
There is no resonance state near the $\Sigma_{c}^{*}N$ threshold.

In the $J^{\pi}=1^{+}$ channel, there are two resonance states, near $\Sigma_{c}N$ and $\Sigma_{c}^{*}N$ thresholds, respectively.
The resonance state near the $\Sigma_{c}N$ threshold has a deeper binding energy and rather broad width.
We find that the $0^{+}$ and $1^{+}$ resonances near the $\Sigma_{c}N$ threshold have different binding energies and widths.
On the other hand the $1^{+}$ resonance near the $\Sigma_{c}^{*}N$ threshold shows rather similar properties as the $0^{+}$ resonance near the $\Sigma_{c}N$ threshold.

In the $J^{\pi}=2^{+}$ channel, there is one resonance state near the $\Sigma_{c}^{*}N$ threshold.
The existence of this resonance state is a distinctive characteristic of two-body charmed baryon - nucleon systems, 
containing the $\Sigma_{c}^{*}N$ channel which has a charmed baryon with spin $S_{Y_{c}}=3/2$.
This resonance state has also a large binding energy and a broad width like the resonance state near the $\Sigma_{c}N$ threshold with $J^{\pi}=1^{+}$.

\begin{figure}[!h]
\begin{tabular}{cc}
\begin{minipage}{0.5\hsize}
\begin{center}
\includegraphics[width=80mm]{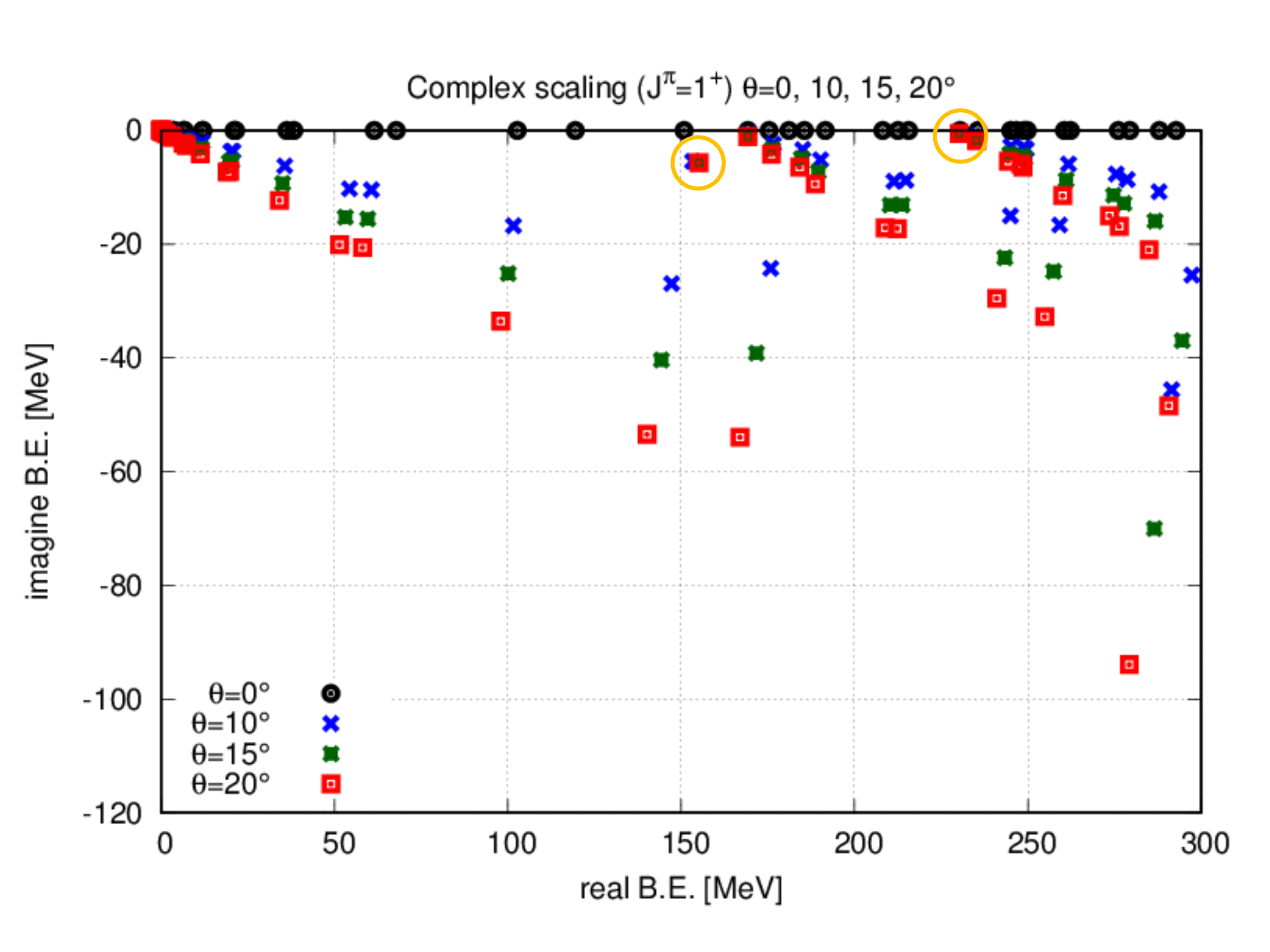}
\caption{\footnotesize{The result of complex scale method in complex plane for $J^{\pi}=1^{+}$} with a-parameter}
\label{gr:res2-8}
\end{center}
\end{minipage}
\begin{minipage}{0.5\hsize}
\begin{center}
\includegraphics[width=80mm]{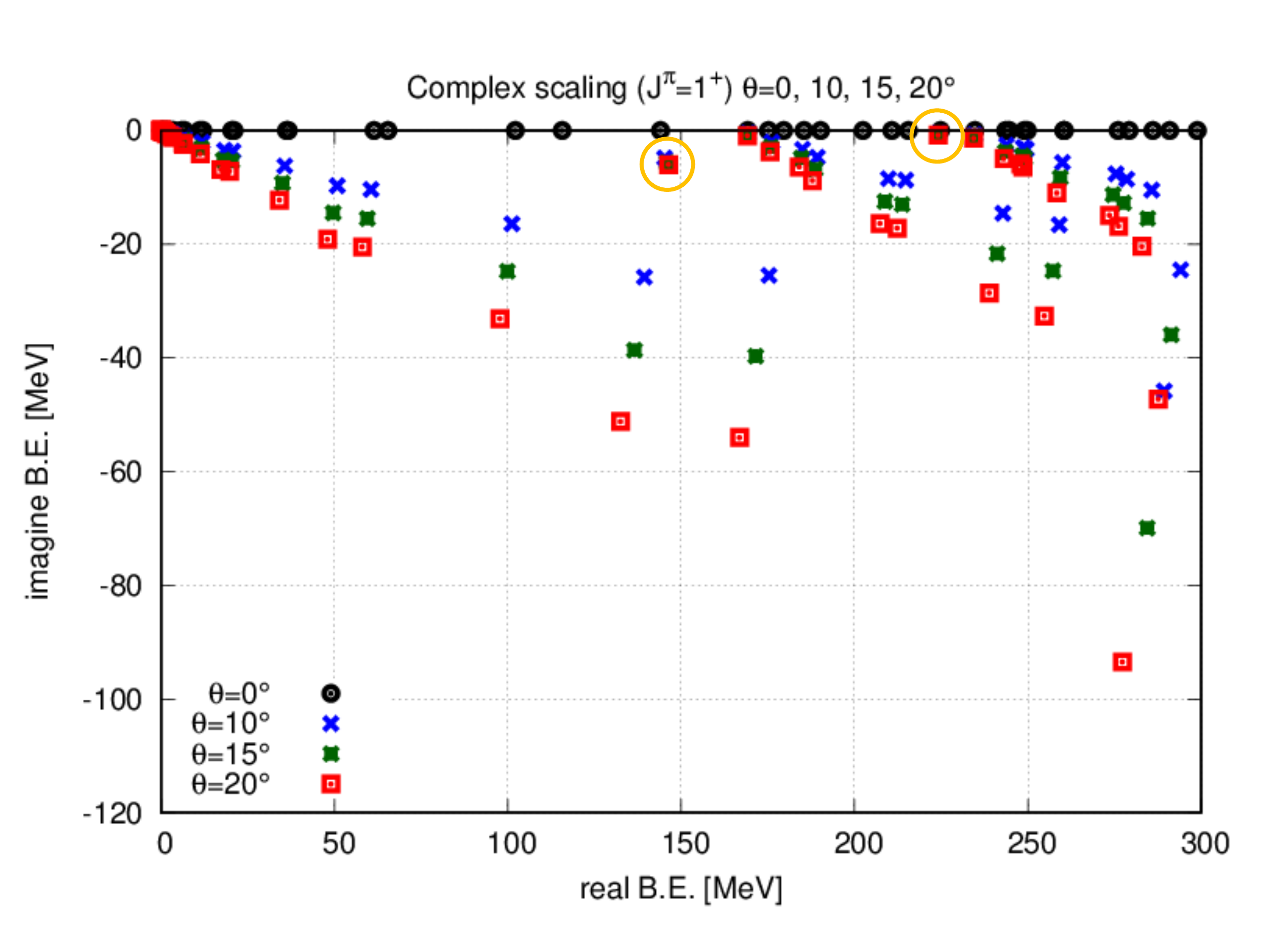}
\caption{\footnotesize{The result of complex scale method in complex plane for $J^{\pi}=1^{+}$} with b-parameter}
\label{gr:res2-9}
\end{center}
\end{minipage}
\end{tabular}
\end{figure}
\begin{figure}[!h]
\begin{center}
\includegraphics[width=80mm]{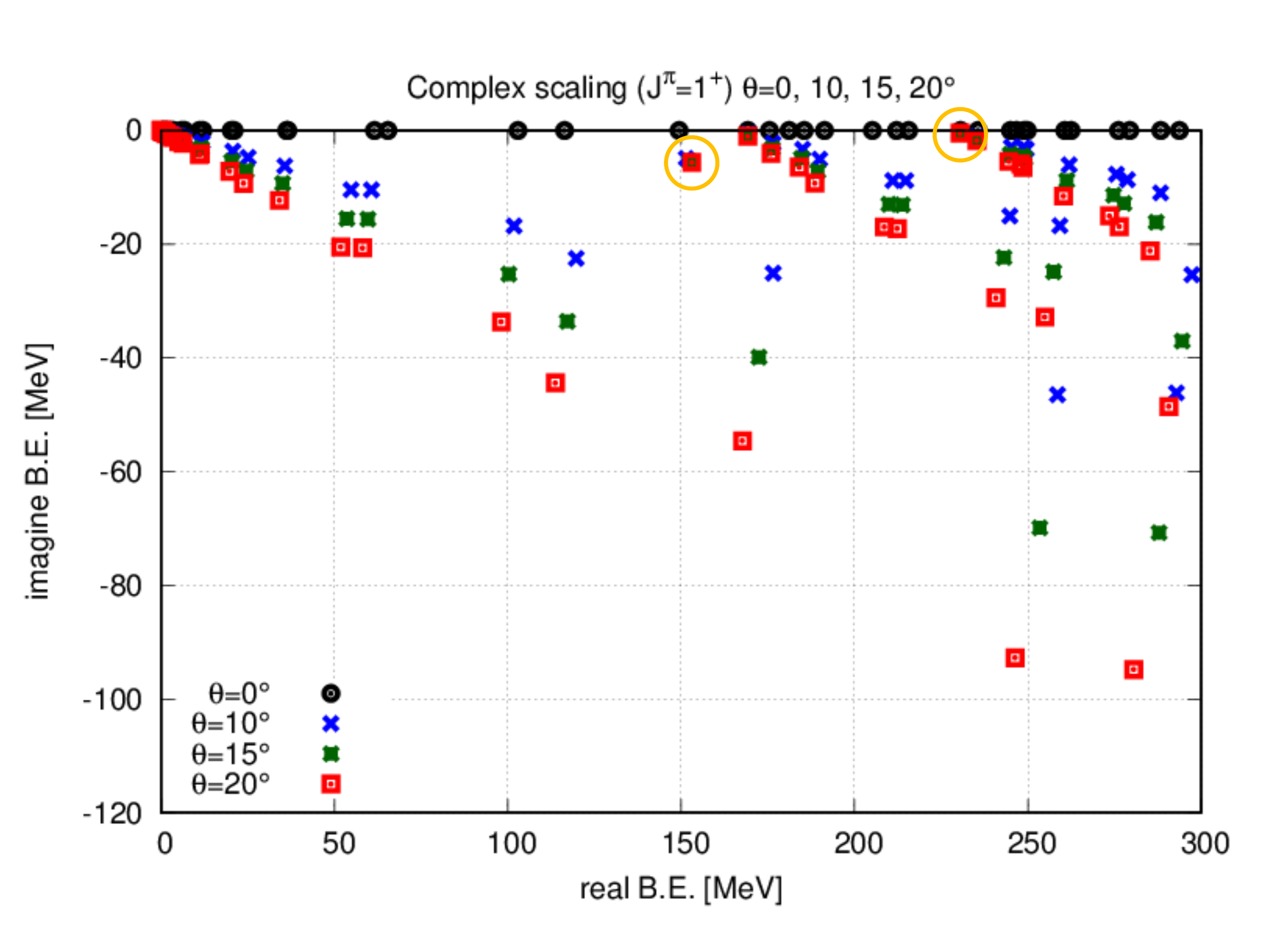}
\caption{\footnotesize{The result of complex scale method in complex plane for $J^{\pi}=1^{+}$} with c-parameter}
\label{gr:res2-10}
\end{center}
\end{figure}

\begin{table}[h]
\centering
\begin{tabular}{r|ccc|c} \hline
 & a-parameter & b-parameter & c-parameter & d-parameter \\ \hline
from $\Lambda_{c}N$ threshold [MeV] & 156 & 147 & 153 & 144  \\
from $\Sigma_{c}N$ threshold [MeV] & -12 & -20 & -14 & -23  \\ \hline
decay width [MeV] & 11 & 12 & 11 & 12 \\ \hline
\end{tabular}
\smallskip
\caption{The resonance states with $Y_{c}N$-CTNN potential with all parameters in $J^{\pi} = 1^{+}$}
\label{tb:res2-c}
\end{table}

We can calculate the resonance states with $Y_{c}N$-CTNN potential not only d-parameter but also a, b, and c-parameter.
These results near the $\Sigma_{c}N$ threshold in $J^{\pi} = 1^{+}$ are shown in Figs. \ref{gr:res2-8}, \ref{gr:res2-9}, and \ref{gr:res2-10} and Table \ref{tb:res2-c}.
According to Table \ref{tb:res2-c}, the resonance energy with b-parameter is larger than that of c-parameter, contrary to the result of binding energy.
In the case of $\Sigma_{c}N$ channel, the repulsive core is weaker than the $\Lambda_{c}N$ channels and it is sensitive to contribution of the $\sigma$ exchange potential.
Therefore, the b-parameter is more attractive than the c parameter because of the strength of the parameter $C_{\sigma}$ related to the $\sigma$ exchange potential.
Although the resonance energies and decay widths are different, the behavior of the resonance state for each state is same in any potential.

\section{Discussion}
\label{sec:res2-5}

\subsection{Comparison of bound states and resonance states}

To investigate the effects of the channel coupling, we calculate the binding energies with partial channel coupling and compare with the resonance results.

\begin{table}[t]
\centering
\begin{tabular}{c|c|cc|cc} \hline
 \multicolumn{2}{c|}{states} & \multicolumn{2}{c|}{B.E.} & \multicolumn{2}{c}{Resonance} \\ \hline
threshold & $J^{\pi}$ & numerical result & (from threshold) & numerical result & (from threshold) \\ \hline
\multirow{2}{*}{$\Sigma_{c}N$} & $J^{\pi}=0^{+}$ & 165 & (-2) & 163 & (-4)  \\
 & $J^{\pi}=1^{+}$ & 153 & (-14) & 144 & (-23) \\ \hline
\multirow{2}{*}{$\Sigma_{c}^{*}N$} & $J^{\pi}=1^{+}$ & 224 & (-8) & 225 & (-7) \\
 & $J^{\pi}=2^{+}$ & 214 & (-18) & 206 & (-25) \\ \hline
\end{tabular}
\smallskip
\caption{Bound states and resonance states}
\label{tb:res2-5}
\end{table}

The binding energies from thresholds of excited states are shown in Table \ref{tb:res2-5}.
Comparing the result of binding energy with that of resonance energy, 
we find that they are roughly consistent. 
Therefore, they are regarded as Feshbach resonances.
Each state resonates with a different bound state.

By comparing the ``binding'' energies and the resonance energies, one finds that
$\Sigma_{c}N$ ($J^{\pi}=0^{+}$) and $\Sigma_{c}^{*}N$ ($J^{\pi}=1^{+}$) are not
affected by the channel coupling effects, while the $\Sigma_{c}N$ ($J^{\pi}=1^{+}$)と$\Sigma_{c}^{*}N$ ($J^{\pi}=2^{+}$) channels have significant effects from the couplings.
Larger widths of the latter states are also consistent with stronger off-diagonal components of the potential.

Fig. 7 shows that the off-diagonal potentials push up the binding energy of the 
$\Sigma_{c}^{*}N (J^{\pi}=1^{+})$, while the effects are attractive in the other channels.

\subsection{Heavy quark spin doublet}

According to the results, we assume that these states can be sorted to 2 groups.
One group has narrow widths and another group has rather broad widths, as shown in Table \ref{tb:res2-5}.
The resonance energy and width of the resonance state near the $\Sigma_{c}^{*}N$ threshold are larger than those of the state near the $\Sigma_{c}N$ threshold.
These properties are not seen in the strangeness sector.

By the way, the heavy quark spin symmetry is important in the charm sector \cite{ref-HQS}.
In particular, $\Sigma_{c}N$ and $\Sigma_{c}^{*}N$ are related to a heavy quark spin doublet and are degenerate in the heavy quark limit.
To investigate these properties, we calculate resonance states with mass degenerate threshold 
in substitution for $\Sigma_{c}N$ and $\Sigma_{c}^{*}N$ thresholds.
The degenerate threshold is obtained from the spin averaged mass, $(m_{\Sigma_{c}N} + 3 m_{\Sigma_{c}^{*}N})/4$.
In order to take the heavy mass limit, we should take into account all the effects of $\Sigma_{c}N$ and $\Sigma_{c}^{*}N$ for potential.
But we change the threshold mass only in this calculation.

\begin{table}[t]
\centering
\begin{tabular}{c||cc|c||cc|c} \hline
 & \multicolumn{3}{c||}{$J^{\pi}=0^{+}$ near $\Sigma_{c}N$} & \multicolumn{3}{c}{$J^{\pi}=1^{+}$ near $\Sigma_{c}^{*}N$} \\ \hline
 & \multicolumn{2}{c|}{resonance} & width & \multicolumn{2}{c|}{resonance} & width  \\ \hline
CTNN-d & 184 & (-5) & 1 & 184 & (-5) & 1  \\ \hline
\end{tabular}
\smallskip
\caption{$Y_{c}N$ resonance states with narrow width for HQ limit (threshold = 188.6 MeV)}
\label{tb:res2-6}
\end{table}
\begin{table}[t]
\centering
\begin{tabular}{c||cc|c||cc|c} \hline
 & \multicolumn{3}{c||}{$J^{\pi}=1^{+}$ near $\Sigma_{c}N$} & \multicolumn{3}{c}{$J^{\pi}=2^{+}$ near $\Sigma_{c}^{*}N$} \\ \hline
 & \multicolumn{2}{c|}{resonance} & width & \multicolumn{2}{c|}{resonance} & width  \\ \hline
CTNN-d & 162 & (-27) & 13 & 161 & (-28) & 14 \\ \hline
\end{tabular}
\smallskip
\caption{$Y_{c}N$ resonance states with wide width for HQ limit (threshold = 188.6 MeV)}
\label{tb:res2-7}
\end{table}

The results are shown in Tables \ref{tb:res2-6} and \ref{tb:res2-7}.
The resonance energies of $\Sigma_{c}N$ increase and those of $\Sigma_{c}^{*}N$ decrease with the degenerate threshold. 
Then we find that these results are almost equal and form heavy quark doublets. 
They are the effects of the heavy quark spin symmetry.

\begin{figure}[!t]
\centering
\includegraphics[width=170mm]{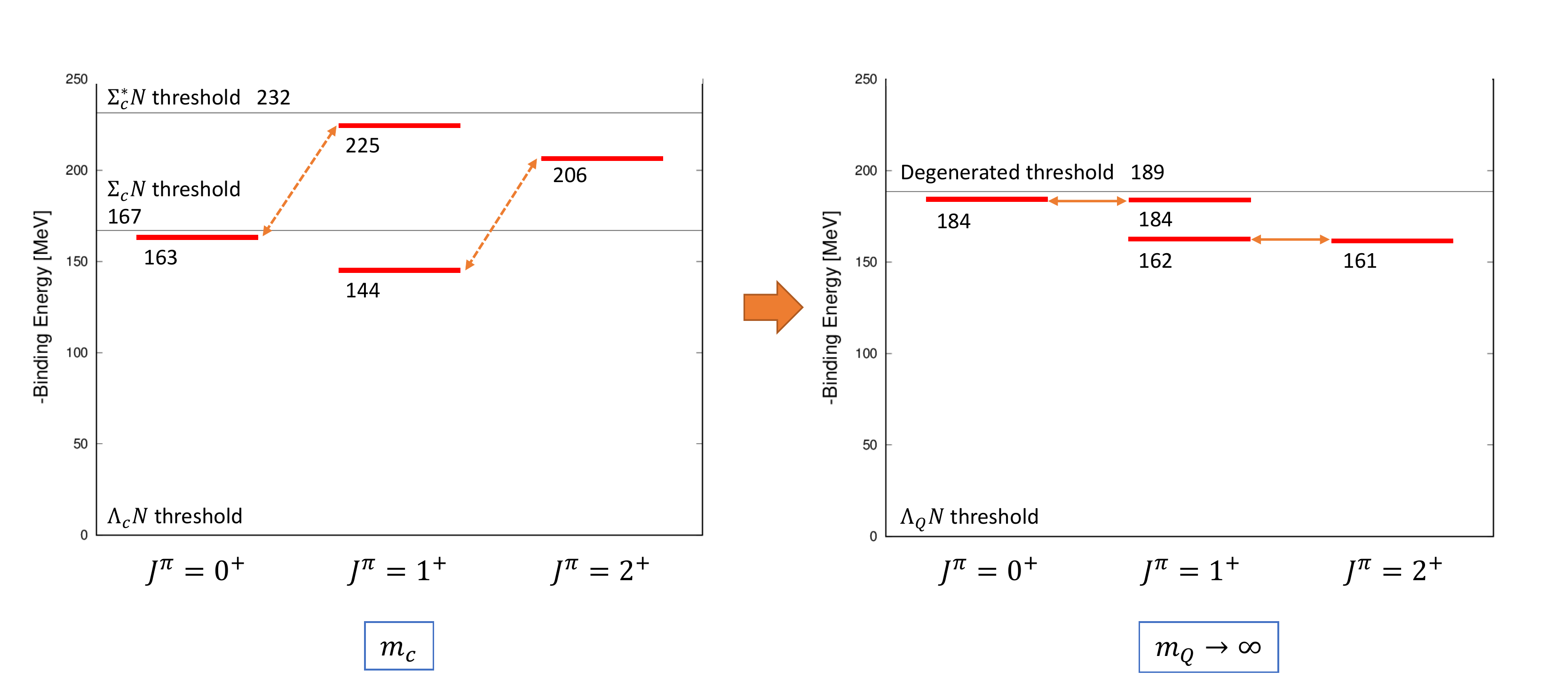}
\caption{The summary of shifting of resonance energies. When $m_{Q} \to \infty$, we substitute  for $\Sigma_{c}N$ and $\Sigma_{c}^{*}N$ thresholds, 
$m_{degenerated} = (m_{\Sigma_{c}} + 3 m_{\Sigma_{c}^{*}N})/4$.}
\label{fig:cc-1}
\end{figure}

Therefore, the results of resonance states and degenerated resonances are summarized as Figure \ref{fig:cc-1}.
In particular, when we adopt the degenerated $\Sigma_{c}N$ and $\Sigma_{c}^{*}N$ thresholds in the heavy quark limit,
the resonance energies make pairs between different total angular momenta. 
These results seem to follow the heavy quark spin symmetry.

\section{Summary}
\label{sec:res2-6}

We have applied the complex scaling method to the $Y_c N$ potential model in order to explore resonance states 
near the $\Sigma_{c}N$ and $\Sigma_{c}^{*}N$ thresholds in the $\Lambda_c N$, $\Sigma_c N$ and $\Sigma_c^* N$ coupled
channel systems.
Four sharp Feshbach-like resonances are located, one for $J^{\pi} = 0^{+}$, two for $1^{+}$, and one for $2^{+}$.
Each of them is shown to correspond to a bound $\Sigma_c N$ or $\Sigma_c^* N$ state when we omit the $\Lambda_c N$
channels.
It is found that the $D$ wave mixings due to the pion-exchange tensor force are significant for the resonances.

Comparing the resonance energies from the thresholds, we observe that they form two groups.
The $0^+$ state at the $\Sigma_c N$ threshold and the $1^+$ state at the $\Sigma_c^*N$ threshold
have ``binding'' energies of less than 10 MeV and very narrow widths  ($\Gamma \sim 1-2$ MeV), 
while the other two, $1^+$ below $\Sigma_c N$ and
$2^+$ have larger ``binding'' energies ($\sim 25$ MeV) from the threshold and larger widths ($\Gamma\sim 10$ MeV).
These behaviors are consistent with the spin doublet states according to the heavy-quark spin symmetry.

The potential model applied here predicts a shallow bound state of $\Lambda_c N$ in $J^{\pi} = 0^{+}$ and $1^{+}$.
It is extremely interesting to find such bound states in experiment.
However, recent lattice calculation predicts less attractive potential so that no two-body $\Lambda_c N$ bound state may exist.
In the present calculation, among the variations of our potential model, the less attractive one also gives the resonance 
states. Thus the search of the $\Sigma_c N$ and/or $\Sigma_c^* N = \Lambda_c+N+\pi$ resonances is also exciting
even if no $\Lambda_c N$ bound state is found.

\section*{Acknowledgment}

S. M. was supported by the RIKEN Junior Research Associate Program.
This work is supported in part by JSPS KAKENHI Grant Nos. 25247036.

\vfill\pagebreak

\newpage

\appendix

\section{Operator calculations for one boson exchange potential with $J^{\pi} = 2^{+}$}
\label{app:obep2}

In this study, we consider the $Y_{c}N$ state with $Y_{c}N$-CTNN potentials for $J^{\pi} = 2^{+}$.
Using these coupled channels, we calculate the expectation values of the matrix elements of 
$\bm{\mathcal{O}}_{spin}$, $\bm{\mathcal{O}}_{ten}$, and $\bm{\mathcal{O}}_{LS}$ for the channels in $I=1/2$, $J^{\pi}=0^{+}$.
We label the relevant channels by $i$ and $j$, and tabulate the $ij$ component of the matrix elements $\left< \bm{\mathcal{O}} \right>_{ij}$.
We present the definitions of the spin dependent operators in the potentials.
$\bm{\mathcal{O}}_{spin}$ is the spin-spin operator between $Y_{c}$ and $N$:
\begin{equation}
\bm{\mathcal{O}}_{spin} = \bm{\mathcal{O}}_1\cdot \bm{\sigma}_2,
\end{equation}
where $\bm{\sigma}_2$ is the Pauli matrix of the nucleon spin, and $\bm{\mathcal{O}}_1$ is the spin operator of the charmed baryon,
\begin{equation}
\bm{\mathcal{O}}_1 = \left\{ 
\begin{array}{cl}
\bm{\sigma}_1 & \rm{for} \ \Lambda_{c} \ \rm{and} \ \Sigma_{c} \\
\bm{\bar{\Sigma}}_1 & \rm{for \ the \ transition \ from} \ \Lambda_{c} \ \rm{and} \ \Sigma_{c} \ \rm{to} \ \Sigma_{c}^{*} \\
\bm{\Sigma}_1 & \rm{for} \ \Sigma_{c}^{*} \\
\end{array} \right.
\end{equation}

The transition spin $\bm{\bar{\Sigma}}$ is defined as $u^{\mu} \equiv \bm{\bar{\Sigma}} \Phi$, where $u^{\mu}$ is the Rarita Schwinger field, 
and $\Phi$ is the spin wave function of $\Sigma_{c}^{*}$, 
\begin{equation}
\Phi(3/2) = \left( \begin{array}{c}1 \\ 0 \\ 0 \\ 0 \\ \end{array} \right), \ 
\Phi(1/2) = \left( \begin{array}{c}0 \\ 1 \\ 0 \\ 0 \\ \end{array} \right), \ 
\Phi(-1/2) = \left( \begin{array}{c}0 \\ 0 \\ 1 \\ 0 \\ \end{array} \right), \ 
\Phi(-3/2) = \left( \begin{array}{c}0 \\ 0 \\ 0 \\ 1 \\ \end{array} \right). \ 
\end{equation}
Then we calculate the transition spin explicitly, 
\begin{equation}
\begin{array}{rcl}
\bm{\bar{\Sigma}}^\dag & = & -\frac{1}{\sqrt{2}}\left( \bm{\bar{\Sigma}}_{x}^\dag + i\bm{\bar{\Sigma}}_{y}^\dag \right) + \frac{1}{\sqrt{2}}\left( \bm{\bar{\Sigma}}_{x}^\dag - i\bm{\bar{\Sigma}}_{y}^\dag \right) + \bm{\bar{\Sigma}}_{z}^\dag \\
 & = & \left(\begin{array}{cccc}1&0&0&0\\0&\sqrt{\frac13}&0&0\end{array}\right)+\left(\begin{array}{cccc}0&\sqrt\frac23&0&0\\0&0&\sqrt\frac23&0\end{array}\right)+\left(\begin{array}{cccc}0&0&\sqrt{\frac13}&0\\0&0&0&1\end{array}\right) \\
\end{array}
\end{equation}
\begin{equation}
\bm{\bar{\Sigma}} \bm{\bar{\Sigma}}^\dag=-I_{4\times4}, \quad 
\bm{\bar{\Sigma}}^\dag \bm{\bar{\Sigma}}=-2I_{2\times2}.
\end{equation}
With $\hat{\bm{e}}(\lambda=+1)=-\frac{1}{\sqrt2}(1,i,0)$, $\hat{\bm{e}}(\lambda=-1)=\frac{1}{\sqrt2}(1,-i,0)$, $\hat{\bm{e}}(\lambda=0)=(0,0,1)$, and $S_{t\mu}^\dag=(0,\vec{S}_t^\dag)$, one has
\begin{eqnarray}
\bm{\bar{\Sigma}}(1,+1)=\left(\begin{array}{cc}1&0\\0&\sqrt{\frac13}\\0&0\\0&0\end{array}\right),\qquad
\bm{\bar{\Sigma}}(1,-1)=\left(\begin{array}{cc}0&0\\0&0\\\sqrt{\frac13}&0\\0&1\end{array}\right),\qquad
\bm{\bar{\Sigma}}(1,0)=\left(\begin{array}{cc}0&0\\\sqrt{\frac23}&0\\0&\sqrt{\frac23}\\0&0\end{array}\right).
\label{eq:trs}
\end{eqnarray}

Next, we define the spin operator of $\Sigma_{c}^{*}$, 
\begin{equation}
\begin{array}{c}
\bm{\Sigma} = -S_{t\mu}^{\dag}\bm{\sigma}S_{t}^{\mu} = (S_{t}^{\dag})^{j}\bm{\sigma}(S_{t})^{j}, \\
\bm{S}(\Sigma_{c}^{*}) = \frac{3}{2}\bm{\Sigma}.\\
\end{array}
\end{equation}
With Eq. (\ref{eq:trs}), the explicit matrices are
\begin{equation}
\begin{array}{c}
\bm{\sigma}_{rs}(1,+1) = -\frac{1}{\sqrt{2}}\left( \bm{\Sigma}_{x} + i\bm{\Sigma}_{y} \right) =-\left( 
\begin{array}{cccc}
0 & \sqrt{\frac{2}{3}} & 0 & 0 \\
0 & 0 & \frac{2\sqrt{2}}{3} & 0 \\
0 & 0 & 0 & \sqrt{\frac{2}{3}} \\
0 & 0 & 0 & 0 \\
\end{array} \right), \\
\bm{\sigma}_{rs}(1,-1) = \frac{1}{\sqrt{2}}\left( \bm{\Sigma}_{x} - i\bm{\Sigma}_{y} \right) = -\left( 
\begin{array}{cccc}
0 & 0 & 0 & 0 \\
\sqrt{\frac{2}{3}} & 0 & 0 & 0 \\
0 & \frac{2\sqrt{2}}{3} & 0 & 0 \\
0 & 0 & \sqrt{\frac{2}{3}} & 0 \\
\end{array} \right), \\
\bm{\sigma}_{rs}(1,0) = \bm{\Sigma}_{z} = -\left( 
\begin{array}{cccc}
1 & 0 & 0 & 0 \\
0 & \frac{1}{3} & 0 & 0 \\
0 & 0 & -\frac{1}{3} & 0 \\
0 & 0 & 0 & -1 \\
\end{array} \right). \\
\end{array}
\end{equation}

The tensor operators can be defined similarly as follows
\begin{eqnarray}
\rm{In} \  \Lambda_{c}  \ \rm{and} \  \Sigma_{c}  \ \rm{channels  \ : \ } \bm{\mathcal{O}}_{ten} = \frac{3(\bm{\sigma}_1 \cdot \bm{r})(\bm{\sigma}_2 \cdot \bm{r})}{r^2} - \bm{\sigma}_1 \cdot \bm{\sigma}_2, \nonumber \\
\rm{In} \  \Lambda_{c} \to \Sigma_{c}^{*}  \ \rm{and}  \ \Sigma_{c} \to \Sigma_{c}^{*}  \ \rm{channels  \ : \ } \bm{\mathcal{O}}_{ten} = \frac{3(\bm{\bar{\Sigma}} \cdot \bm{r})(\bm{\sigma}_2 \cdot \bm{r})}{r^2} - \bm{\bar{\Sigma}} \cdot \bm{\sigma}_2, \nonumber \\
\rm{In} \  \Sigma_{c}^{*}   \ \rm{diagonal  \ channels  \ : \ } \bm{\mathcal{O}}_{ten} = \frac{3(\bm{\Sigma} \cdot \bm{r})(\bm{\sigma}_2 \cdot \bm{r})}{r^2}-\bm{\Sigma} \cdot \bm{\sigma}_2,
\end{eqnarray}

The spin-orbit operator $\bm{\mathcal{O}}_{LS}$ is defined as
\begin{equation}
\bm{\mathcal{O}}_{LS} = \bm{L}\cdot \bm{\sigma}_2.
\end{equation}
$\bm{L}\cdot \bm{\mathcal{O}}_1$ is not included in the potential for this calculation.

\subsection{$I=\frac12$, $J^{\pi}=2^+$ coupled system}

The 8 channel coupling matrix elements are given in Tables \ref{tb:op7}-\ref{tb:op9} for $I=1/2$, $J^{\pi}=2^{+}$.

\begin{table}[htb]
\begin{center}
\scalebox{0.8}{
\begin{tabular}{c|cccccccc}\hline
\backslashbox{i}{j} & $\Lambda_cN(^1D_2)$&$\Sigma_cN(^1D_2)$&$\Lambda_cN(^3D_2)$&$\Sigma_cN(^3D_2)$&$\Sigma_c^*N(^3D_2)$&$\Sigma_c^*N(^5S_2)$&$\Sigma_c^*N(^5D_2)$&$\Sigma_c^*N(^5G_2)$ \\ \hline
$\Lambda_cN(^1D_2)$ & -3 & -3 & 0 & 0 & 0 & 0 & 0 & 0 \\
$\Sigma_cN(^1D_2)$  & -3 & -3 & 0 & 0 & 0 & 0 & 0 & 0 \\
$\Lambda_cN(^3D_2)$ & 0 & 0 & 1 & 1 & $-\sqrt\frac83$ & 0 & 0 & 0 \\
$\Sigma_cN(^3D_2)$  & 0 & 0 & 1 & 1 & $-\sqrt\frac83$ & 0 & 0 & 0 \\
$\Sigma_c^*N(^3D_2)$& 0 & 0 & $-\sqrt\frac83$ & $-\sqrt\frac83$ & $-\frac53$ & 0 & 0 & 0 \\
$\Sigma_c^*N(^5S_2)$& 0 & 0 & 0 & 0 & 0 & 1 & 0 & 0 \\
$\Sigma_c^*N(^5D_2)$& 0 & 0 & 0 & 0 & 0 & 0 & 1 & 0 \\
$\Sigma_c^*N(^5G_2)$& 0 & 0 & 0 & 0 & 0 & 0 & 0 & 1 \\ \hline
\end{tabular}
}
\caption{The matrix elements of the spin-spin operators $\left< \bm{\mathcal{O}}_{spin} \right>_{ij}$ for the $I=\frac12$, $J^{\pi}=2^+$ coupled system.}
\label{tb:op7}
\end{center}
\end{table}

\begin{table}[htb]
\begin{center}
\scalebox{0.8}{
\begin{tabular}{c|cccccccc}\hline
\backslashbox{i}{j} & $\Lambda_cN(^1D_2)$&$\Sigma_cN(^1D_2)$&$\Lambda_cN(^3D_2)$&$\Sigma_cN(^3D_2)$&$\Sigma_c^*N(^3D_2)$&$\Sigma_c^*N(^5S_2)$&$\Sigma_c^*N(^5D_2)$&$\Sigma_c^*N(^5G_2)$ \\\hline
$\Lambda_cN(^1D_2)$ & 0 & 0 & 0 & 0 & 0 & $-\sqrt\frac65$ & $\frac{2\sqrt{21}}{7}$ & $-\sqrt{\frac{108}{35}}$ \\
$\Sigma_cN(^1D_2)$  & 0 & 0 & 0 & 0 & 0 & $-\sqrt\frac65$ & $\frac{2\sqrt{21}}{7}$ & $-\sqrt{\frac{108}{35}}$ \\
$\Lambda_cN(^3D_2)$ & 0 & 0 & 2 & 2 & $\sqrt\frac16$ & $-\frac{3}{\sqrt{5}}$ & $\frac{3}{\sqrt{14}}$ & $\sqrt{\frac{72}{35}}$ \\
$\Sigma_cN(^3D_2)$  & 0 & 0 & 2 & 2 & $\sqrt\frac16$ & $-\frac{3}{\sqrt{5}}$ & $\frac{3}{\sqrt{14}}$ & $\sqrt{\frac{72}{35}}$ \\
$\Sigma_c^*N(^3D_2)$& 0 & 0 & $\sqrt\frac16$ & $\sqrt\frac16$ & $-\frac13$ & $\sqrt{\frac65}$ & $-\sqrt{\frac37}$ & $-\sqrt{\frac{48}{35}}$ \\
$\Sigma_c^*N(^5S_2)$& $-\sqrt\frac65$ & $-\sqrt\frac65$ & $-\frac{3}{\sqrt{5}}$ & $-\frac{3}{\sqrt{5}}$ & $\sqrt{\frac65}$ & 0  & $\sqrt{\frac{14}{5}}$ & 0\\
$\Sigma_c^*N(^5D_2)$& $\frac{2\sqrt{21}}{7}$ & $\frac{2\sqrt{21}}{7}$ & $\frac{3}{\sqrt{14}}$ & $\frac{3}{\sqrt{14}}$ & $-\sqrt{\frac{3}{7}}$ & $\sqrt{\frac{14}{5}}$ & $\frac{3}{7}$ & $\frac{12\sqrt5}{35}$ \\
$\Sigma_c^*N(^5G_2)$& $-\sqrt{\frac{108}{35}}$ & $-\sqrt{\frac{108}{35}}$ & $\sqrt{\frac{72}{35}}$ & $\sqrt{\frac{72}{35}}$ & $-\sqrt{\frac{48}{35}}$ & 0 & $\frac{12\sqrt5}{35}$ & $-\frac{10}{7}$ \\ \hline
\end{tabular}
}
\caption{The matrix elements of the tensor operators $\left< \bm{\mathcal{O}}_{ten} \right>_{ij}$ for the $I=\frac12$, $J^{\pi}=2^+$ coupled system.}
\label{tb:op8}
\end{center}
\end{table}

\begin{table}[htb]
\begin{center}
\scalebox{0.8}
{\begin{tabular}{c|cccccccc}\hline
\backslashbox{i}{j} & $\Lambda_cN(^1D_2)$&$\Sigma_cN(^1D_2)$&$\Lambda_cN(^3D_2)$&$\Sigma_cN(^3D_2)$&$\Sigma_c^*N(^3D_2)$&$\Sigma_c^*N(^5S_2)$&$\Sigma_c^*N(^5D_2)$&$\Sigma_c^*N(^5G_2)$ \\\hline
$\Lambda_cN(^1D_2)$ & 0 & 0 & $-\sqrt6$ & $-\sqrt6$ & 0 & 0 & 0 & 0 \\
$\Sigma_cN(^1D_2)$  & 0 & 0 & $-\sqrt6$ & $-\sqrt6$ & 0 & 0 & 0 & 0 \\
$\Lambda_cN(^3D_2)$ & $-\sqrt6$ & $-\sqrt6$ & -1 & -1 & 0 & 0 & 0 & 0 \\
$\Sigma_cN(^3D_2)$  & $-\sqrt6$ & $-\sqrt6$ & -1 & -1 & 0 & 0 & 0 & 0 \\
$\Sigma_c^*N(^3D_2)$& 0 & 0 & 0 & 0 &$\frac12$ & 0 & $-\frac{\sqrt{21}}{2}$ & 0 \\
$\Sigma_c^*N(^5S_2)$& 0 & 0 & 0 & 0 & 0 & 0 & 0 & 0 \\
$\Sigma_c^*N(^5D_2)$& 0 & 0 & 0 & 0 & $-\frac{\sqrt{21}}{2}$ & 0 & -$\frac{3}{2}$ & 0 \\
$\Sigma_c^*N(^5G_2)$& 0 & 0 & 0 & 0 & 0 & 0 & 0 & -5 \\ \hline
\end{tabular}
}
\caption{The matrix elements of the orbital-spin operators $\left< \bm{\mathcal{O}}_{LS} \right>_{ij}$ for the $I=\frac12$, $J^{\pi}=2^+$ coupled system.}
\label{tb:op9}
\end{center}
\end{table}

\end{document}